\documentclass{PoS}

\title{
Supersymmetric Z$'$ decays at the LHC
}

\ShortTitle{
Supersymmetric Z$'$ decays at the LHC
}

\author{
\speaker{Gennaro Corcella}%
        \\
        INFN, Laboratori Nazionali di Frascati\\
        Via E.~Fermi 40, I-00044 Frascati (RM), Italy\\
        E-mail: \email{gennaro.corcella@lnf.infn.it}
}

\abstract{
Searching for $Z'$ bosons, predicted in GUT-inspired U(1)$'$ gauge models
and in the Sequential Standard Model, 
is one of the main challenges of the experiments carried out
at the Large Hadron Collider.
Such searches have so far focused on high-mass
dilepton pairs, assuming that the $Z'$ has only Standard Model decay
modes, and have set mass exclusion limits around 2.5-3 TeV. 
In this talk, I investigate 
supersymmetric $Z'$ decays at 14 TeV LHC,
extending the MSSM in such a way to accommodate extra heavy gauge
bosons. In particular, I study $Z'$ decays into pairs of sleptons,
charginos and neutralinos, leading to final states with leptons
and missing energy, and present results for few reference 
points of the parameter space, consistent with a SM-like Higgs boson
with a mass around 125 GeV. I also discuss the feasibility to
search for Dark Matter candidates, by analysing $Z'$ decays into
the lightest MSSM neutralinos.
}

\FullConference{
18th International Conference From the Planck Scale to the
Electroweak Scale \\
                 25-29 May 2015\\
                 Ioannina, Greece 
}

\begin{document}
\section{Introduction}
Heavy neutral gauge bosons $Z'$ are predicted in extensions of
the Standard Model (SM), based on U(1)$'$ gauge symmetries, typically inspired
by Grand Unification Theories (see, e.g., \cite{langa} for a 
review). Moreover, they are present in the so-called
Sequential Standard Model (SSM), 
wherein $Z'$ and possibly $W'$ bosons have the same couplings
to fermions as the $Z$ and $W$.

The LHC experiments have so far searched for the $Z'$ assuming that
it decays into SM channels, focusing on high-mass
electron or muon pairs.
The absence of this signal has led 
the ATLAS Collaboration to set the limits 
$m_{Z'}>2.90$~TeV in the SSM and $m_{Z'}>2.51-2.62$~TeV in
U(1)$'$ models \cite{atlas}, and CMS to set $m_{Z'}>2.90$~TeV (SSM) and   
$m_{Z'}>2.57$~TeV (GUTs) \cite{cms}.
Hereafter, I study possible decays of the $Z'$ beyond the
Standard Model (BSM), taking particular care about 
supersymmetric final states, within the Minimal Supersymmetric
Standard Model (MSSM). 

Supersymmetric $Z'$ decays were first considered in 
\cite{gherghetta} and lately reconsidered in \cite{corgen,cor1,cor},
in both GUT-inspired and SSM frameworks.
Although the SM modes still dominate, 
the opening of BSM decay channels decreases the 
branching ratio into dilepton pairs and therefore modifies the
LHC exclusion limits.
Reference~\cite{cor1}, using the same representative point of the MSSM 
parameter space as in \cite{corgen},
found that the LHC exclusion limits decrease by an amount
$\Delta m_{Z'}\simeq$~150-300~GeV, once including supersymmetric
decays. However, Refs.~\cite{corgen,cor1} chose a reference point
which is consistent with the present limits on
supersymmetry, but not with the observed Higgs mass,
around 125 GeV.
Reference~\cite{cor} extended the investigation in \cite{corgen} and,
by fully including one-loop corrections to the Higgs mass, managed
to identify points of the parameter space which are consistent
with both supersymmetry searches and Higgs-boson properties. 

From the viewpoint of supersymmetry, the production of
sparticles in $Z'$ decays has
the advantage that their invariant mass is fixed 
by the heavy boson mass: therefore, if one had to discover a
$Z'$, it would be a cleaner channel to search for supersymmetry,
with respect to direct production in 
$q\bar q$ or $gg$ annihilation.
Moreover, $Z'$ decays into the lightest neutralinos
are interesting processes to search for Dark Matter
candidates, since they lead to mono-jet and mono-photon final states
(see \cite{belang} for a recent study on
Dark Matter at the LHC,
within the U(1)$'$ extension of the MSSM).

In this talk, I shall highlight the main results of Ref.~\cite{cor},
investigating the feasibility to search for supersymmetry
in $Z'$ decays at 14 TeV in U(1)$'$ models.

\section{Theoretical framework}
U(1)$'$ gauge groups and $Z'$ bosons arise in the 
framework of a rank-6 Grand Unification group ${\rm E}_6$.
The $Z'_\psi$ is associated with U(1)$'_\psi$, coming from the
breaking of ${\rm E}_6$ into SO(10):
\begin{equation}\label{upsi}
{\rm E}_6\to {\rm SO}(10)\times {\rm U}(1)'_\psi.
\end{equation}
The subsequent breaking of SO(10) into SU(5) leads to the U(1)$'_\chi$ 
group and $Z'_\chi$ bosons:
\begin{equation}
{\rm SO}(10)\to {\rm SU}(5)\times {\rm U}(1)'_\chi.
\end{equation}
A generic $Z'$ is then a mixture of
$Z'_\psi$ and $Z'_\chi$ bosons, with a mixing angle $\theta$:
\begin{equation}\label{ztheta}
Z'(\theta)=Z'_\psi\cos\theta-Z'_\chi\sin\theta.
\end{equation}
Another scenario, typical of superstring theories, 
is a direct breaking of E$_6$ in the SM and a U(1)$'_\eta$ group, leading
to a $Z'_\eta$ boson, with a mixing angle 
$\theta=\arccos\sqrt{5/8}$:
\begin{equation}\label{ueta}
{\rm E}_6\to {\rm SM}\times U(1)'_\eta.
\end{equation}
As in \cite{cor}, in this talk I shall mostly concentrate the analysis on the
the $Z'_\psi$ and $Z'_\eta$ models, as they are the most interesting
in the supersymmetric extension of the SM.

In fact, the MSSM gets a few novel features, due to
presence of the $Z'$ boson. 
In addition to the scalar Higgs doublets $H_d$
and $H_u$, an extra neutral singlet $S$ is necessary
to break U(1)$'$ and give mass to the $Z'$.
After electroweak symmetry breaking, the Higgs sector consists of
one pseudoscalar $A$ and three scalars $h$, $H$ and $H'$, where
$H'$ is the new boson, due to the extra U(1)$'$.
In the gaugino sector,  
two more neutralinos are present, associated with the
supersymmetric partners of $Z'$ and $H'$.

Furthermore, as thoroughly debated in \cite{gherghetta}, the U(1)$'$ group
leads to extra D- and F-term
contributions to the sfermion masses. In particular, 
the D-term, depending on the sfermion and Higgs
U(1)$'$ charges, can be large and negative, 
so that even the sfermion squared masses can become negative
and thus unphysical (see \cite{corgen} for a few examples of unphysical
configurations).

Hereafter, the $Z'$ mass
will always be set to 
$m_{Z'}=2~{\rm TeV}$ and the coupling
constants of U(1)$'$ and U(1)$_{\rm Y}$, i.e.
$g'$ and $g_1$, are assumed to be proportional, 
as often happens in GUTs:
\begin{equation}
g'=\sqrt{\frac{5}{3}}g_1.
\end{equation}
The supersymmetric parameters $\tan\beta=v_u/v_d$, $\mu$,
$M_1$, and $M'$, where $M_1$ and $M'$ are the soft masses of the
gauginos $\tilde B$ and $\tilde B'$, are fixed as follows:
\begin{equation}
M_1=400~{\rm GeV},\ M'=1~{\rm TeV},\ \tan\beta=30,
\ \mu=200~{\rm GeV}.
\label{mubeta}
\end{equation}
Given $M_1$, the wino mass $M_2$ can be obtained through
$M_2=(3/5)\cot^2\theta_W\simeq 827$~GeV, $\theta_W$ being the Weinberg
angle. 
As for the trilinear couplings $A_{q,\ell}$  of squarks and 
sleptons with the Higgs in the soft supersymmetric Lagrangian
and $A_\lambda$, the soft trilinear coupling of the 
the three Higgs fields ($H_u$, $H_d$ and $S$), they
are fixed to the same value:
\begin{equation}
A_q=A_\ell=A_\lambda= 4~{\rm TeV}.
\label{aaa}
\end{equation}

\section{Results}
In this section I present a few results for the models U(1)$'_\psi$ and
U(1)$'_\eta$.
In each scenario, a point of the parameter space
is chosen, in such a way that at least one decay mode into
supersymmetric particles is substantial and leads to an
observable signal at the LHC.
Besides, in order to suppress QCD backgrounds, 
particular care will be taken about leptonic final states.

\subsection
{Phenomenology - U$(1)'_\psi$ model}

The $Z'_\psi$ model corresponds to a mixing angle $\theta=0$
in Eq.~(\ref{ztheta}).
Following \cite{cor1}, the $Z'_\psi$ mass is set to $m_{Z'}=2$~TeV,
the MSSM parameters as in Eqs.~(\ref{mubeta}) and (\ref{aaa}), 
whereas the soft sfermion masses at the $Z'_\psi$ mass scale are given by:
\begin{equation}
\label{sfer1}
m_{\tilde\ell}^0=m_{\tilde\nu_\ell}^0=1.2~{\rm TeV}\ ,\ 
m^0_{\tilde q}=5.5~{\rm TeV}.
\end{equation}
The squark and slepton masses, obtained by summing to the numbers in 
Eq.~(\ref{sfer1}) the D- and
F-terms, computed by means of the SARAH \cite{sarah} and SPheno 
\cite{spheno} codes, 
are quoted in 
Tables~\ref{tabmassq} and \ref{tabmassl}.
The notation $q_{1,2}$ 
and $\ell_{1,2}$ refers to the
mass eigenstates, determined from the weak states
$q_{L,R}$ and $\ell_{L,R}$,
after diagonalizing the mass mixing matrices.
Tables~\ref{tabmassh} and \ref{tabmasscn} contain the mass spectra
of Higgs bosons and gauginos (charginos and neutralinos), respectively.
In the Higgs sector, the lightest $h$ has a mass compatible
with the SM Higgs, $H$ is roughly as heavy as the $Z'$, the novel scalar
$H'$, the pseudoscalar $A$ and the charged $H^\pm$ are all above
4 TeV and therefore too heavy to contribute to the decay width
of a 2 TeV $Z'_\psi$. As for the gauginos, with the exception of
the heaviest neutralino $\tilde\chi^0_6$, they are lighter than the
$Z'_\psi$.
\begin{table}[htp]
\caption{Squark masses in GeV the reference point of  U$\protect(1)'_\psi$.}
\label{tabmassq}
\begin{center}
\small
\begin{tabular}{|c|c|c|c|c|c|}
\hline
$m_{\tilde d_1}$ &   $m_{\tilde u_1}$ &  $m_{\tilde s_1}$ &  $m_{\tilde c_1}$ &  
$m_{\tilde b_1}$ & $m_{\tilde t_1}$\\ 
\hline
 5609.8 & 5609.4  & 5609.9 & 5609.5 & 2321.7 & 2397.2 \\
\hline
$m_{\tilde d_2}$ &   $m_{\tilde u_2}$ &  $m_{\tilde s_2}$ &  $m_{\tilde c_2}$ &  
$m_{\tilde b_2}$ & $m_{\tilde t_2}$\\ 
\hline
 5504.9 & 5508.7  & 5504.9 & 5508.7 & 2119.6 & 2036.3 \\
\hline\end{tabular}
\end{center}
\end{table}\par
\begin{table}[htp]
\caption{Masses of charged sleptons ($\protect\ell=e,\mu$)
and sneutrinos in the reference point of the U(1)$'_\psi$ model.}
\label{tabmassl}
\begin{center}
\small
\begin{tabular}{|c|c|c|c|c|c|c|c|}
\hline
$m_{\tilde \ell_1}$ &   $m_{\tilde \ell_2}$ &  $m_{\tilde\tau_1}$ & $m_{\tilde\tau_2}$ &
$m_{\tilde \nu_{\ell,1}}$ &  $m_{\tilde \nu_{\ell,2}}$ &
$m_{\tilde \nu_{\tau,1}}$ &  $m_{\tilde \nu_{\tau,2}}$ \\ 
\hline
 1392.4 & 953.0  & 1398.9 & 971.1 & 1389.8  & 961.5 & 1395.9 & 961.5\\ 
\hline\end{tabular}
\end{center}
\end{table}
\begin{table}[htp]
\caption{Masses of neutral and charged Higgs bosons.}
\label{tabmassh}
\begin{center}
\small
\begin{tabular}{|c|c|c|c|c|}
\hline
$m_h$ &   $m_H$&  $m_{H'}$ &  $m_A$ & $m_{H^\pm}$\\ 
\hline
 125.0 &  1989.7 & 4225.0  & 4225.0 & 4335.6 \\ 
\hline\end{tabular}
\end{center}
\end{table}
\begin{table}[htp]
\caption{Masses of charginos 
and neutralinos in the reference point of
the U(1)$'_\psi$ model.}
\label{tabmasscn}
\begin{center}
\small
\begin{tabular}{|c|c|c|c|c|c|c|c|}
\hline
$m_{\tilde\chi_1^+}$ &   $m_{\tilde\chi_2^+}$ & $m_{\tilde\chi_1^0}$ &   $m_{\tilde\chi_2^0}$ 
& $m_{\tilde\chi_3^0}$ &   $m_{\tilde\chi_4^0}$ & $m_{\tilde\chi_5^0}$ &   $m_{\tilde\chi_6^0}$  \\ 
\hline
 204.8 & 889.1 & 197.2  & 210.7 & 408.8 & 647.9 & 889.0 & 6193.5 \\ 
\hline\end{tabular}
\end{center}
\end{table}\par
\begin{table}[htp]
\caption{$\protect Z'_\psi$ decay rates for $\protect m_Z'=2$~TeV.}
\label{tabbrpsi}
\begin{center}
\small
\begin{tabular}{|c|c|}
\hline
Final State & $Z'_\psi$ Branching ratio (\%) \\
\hline
$\tilde\chi_1^+\chi_1^-$ & 10.2 \\
\hline
$\tilde\chi_1^0\tilde\chi_1^0$ & 4.9 \\
\hline
$\tilde\chi_1^0\tilde\chi_3^0$ & 0.2 \\
\hline
$\tilde\chi_2^0\tilde\chi_2^0$ & 5.1 \\
\hline
$\tilde\chi_4^0\tilde\chi_4^0$ & 8.0  \\
\hline
$hZ$ & 1.4\\
\hline
$W^+W^-$ & 2.9\\
\hline
$\sum_i d_i\bar d_i$ & 25.1\\
\hline
$\sum_i u_i\bar u_i$ & 25.0 \\
\hline
$\sum_i \nu_i\bar \nu_i$ & 8.3 \\
\hline
$\sum_i \ell^+_i\ell^-_i$ & 8.3 \\
\hline\end{tabular}
\end{center}
\end{table}
\begin{table}[htp]
\caption{Chargino $\protect\tilde\chi_1^+$ 
decay rates in the reference point of
the $\protect Z'_\psi$ model.}
\label{tabbrch}
\begin{center}
\small
\begin{tabular}{|c|c|}
\hline
Final State & $\chi_1^+$ branching ratio (\%) \\
\hline
$\tilde\chi^0_1\  u\bar d$ & 34.3 \\
\hline
$\tilde\chi^0_1\  u\bar c$ & 1.8 \\
\hline
$\tilde\chi_1^0\  c\bar d$ & 1.6 \\
\hline
$\tilde\chi_1^0\  c\bar s$ & 29.3 \\
\hline
$\tilde\chi_1^0\  e^+\nu_e$ & 12.0 \\
\hline
$\tilde\chi_1^0\  \mu^+\nu_\mu$ & 12.0 \\
\hline
$\tilde\chi_1^0\  \tau^+\nu_\tau$ & 8.9 \\
\hline\end{tabular}
\end{center}
\end{table}\par
The $Z'_\psi$ branching ratios are
quoted in Table~\ref{tabbrpsi}, 
neglecting rates which are below 0.1\%.
The overall branching ratio into supersymmetric final states is
28.3\%; the rate into chargino pairs 
$\tilde\chi^+_1\tilde\chi^-_1$ accounts
for about 10\%, while
decays into pairs of the lightest neutralinos, i.e. 
$\tilde\chi^0_1\tilde\chi^0_1$, possibly relevant 
for the searches for Dark Matter,
for almost 5\%.

Investigating in more detail the $Z'_\psi\to \tilde\chi_1^+\tilde\chi_1^-$
process, subsequent $\tilde\chi_1^\pm$ decays may lead to final states with
charged leptons and missing energy, as in the following process
($\ell=\mu,e$):
\begin{equation}\label{zpc1c1}
pp\to Z'_\psi\to\tilde\chi_1^+\tilde\chi_1^-\to
(\tilde\chi_1^0\ell^+\nu_\ell)(\tilde\chi_1^0\ell^-\bar\nu_\ell).
\end{equation} 
The decay rates of the charginos are reported in Table~\ref{tabbrch};
the cross section of the process in Eq.~(\ref{zpc1c1}),
computed by MadGraph \cite{madgraph}, reads:
$\sigma (pp\to Z'_\psi\to \ell^+\ell^- +\rm{MET})\simeq 7.9\times 10^{-4}$~pb
at 14 TeV.
One may therefore expect about 80 events 
for a luminosity ${\cal L}\simeq 100~{\rm fb}^{-1}$,
almost 240 at 300 fb$^{-1}$.

In the following, I
will present some leptonic distributions
and compare them with direct decays, $pp\to Z'_\psi\to \ell^+\ell^-$,
and direct chargino production, i.e.
\begin{equation}\label{c1c1dir}
pp\to \tilde\chi_1^+\tilde\chi_1^-\to
(\tilde\chi_1^0\ell^+\nu_\ell)(\tilde\chi_1^0\ell^-\bar\nu_\ell).
\end{equation}
The cross section of (\ref{c1c1dir})
is roughly $\sigma\simeq 1.15\times 10^{-2}$~pb.
Furthermore, in (\ref{zpc1c1}) 
the $\tilde\chi_1^+\tilde\chi_1^-$ invariant mass reproduces
$m_{Z'}$, while in (\ref{c1c1dir}) the charginos
do not have this constraint and can be rather soft. 

\begin{figure}[t!]
\centerline{\resizebox{0.42\textwidth}{!}
{\includegraphics{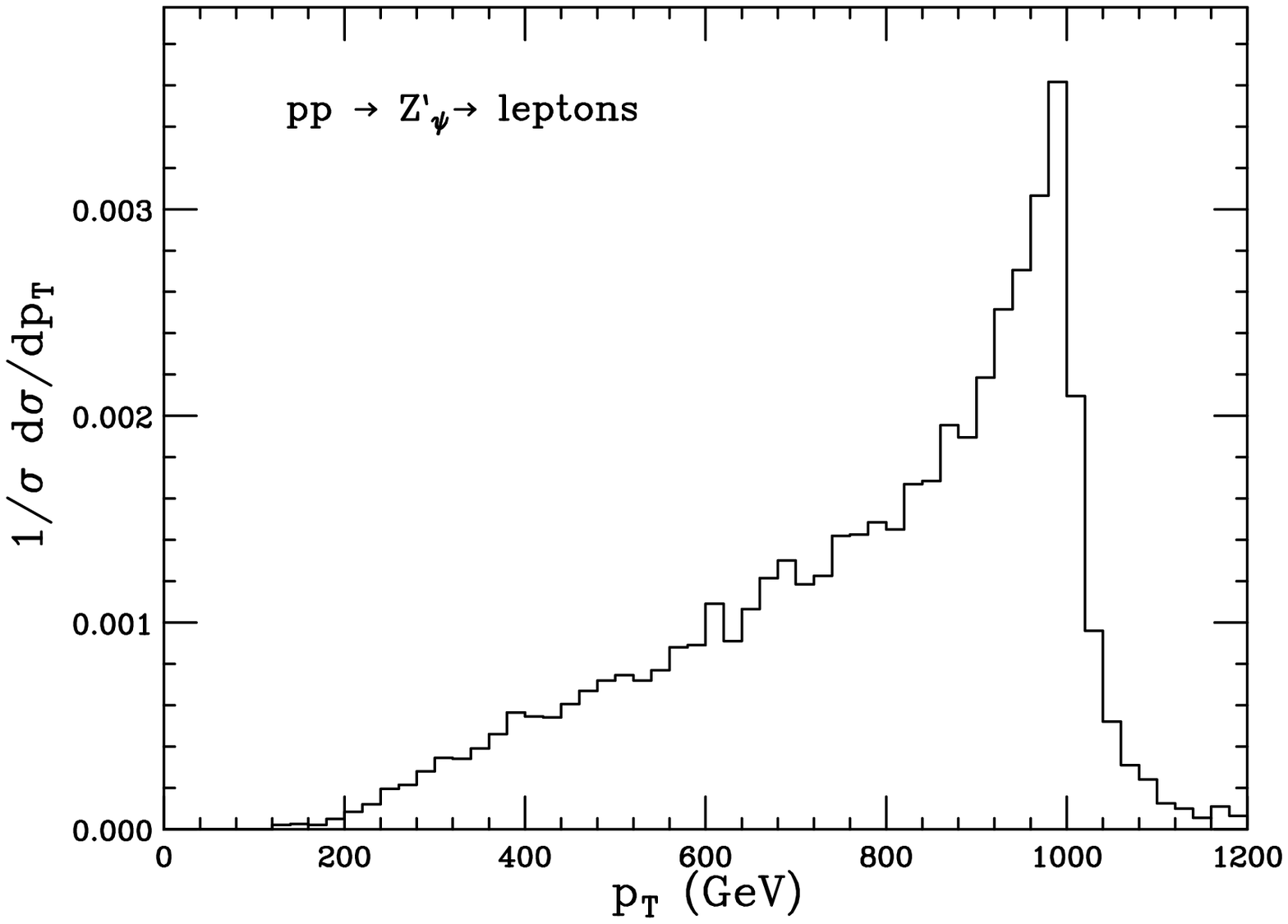}}%
\hfill%
\resizebox{0.42\textwidth}{!}{\includegraphics{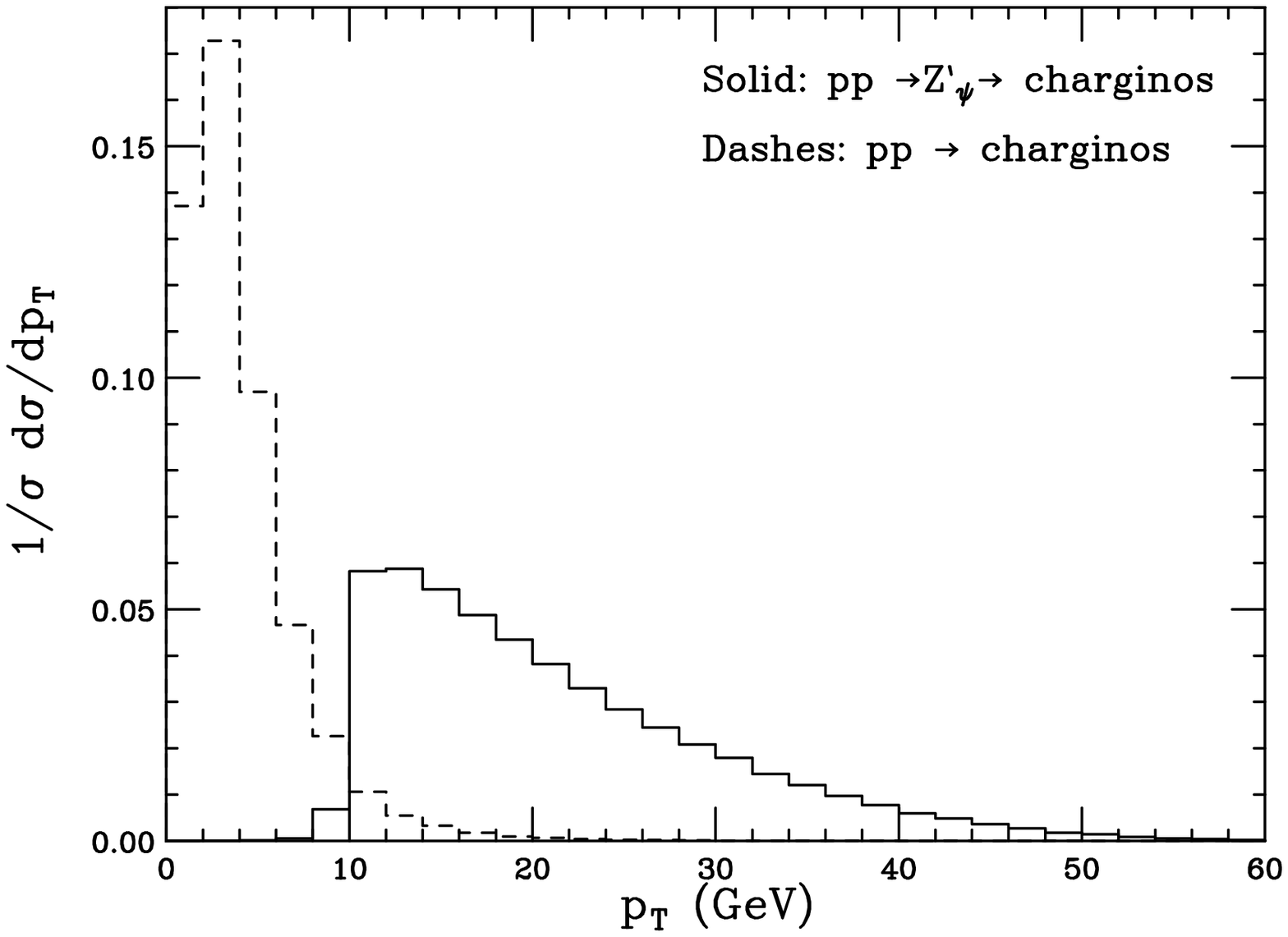}}}
\caption{Lepton transverse momentum for the $Z'_\psi$ model at $\sqrt{s}=14$~TeV
and $\protect 
m_{Z'}=2$~TeV, for a direct $\protect Z'_\psi\to \ell^+\ell^-$ decay (left) and 
chains initiated by $\protect Z'_\psi\to\tilde\chi^+_1\tilde\chi^-_1$ or direct 
chargino production processes (right).}
\label{zpsipt}
\end{figure}

Figure~\ref{zpsipt} presents the transverse momentum of leptons
produced in all three processes, according to
MadGraph interfaced to HERWIG \cite{herwig} 
for parton showers and hadronization.
In direct $Z'_\psi\to\ell^+\ell^-$ events, 
the two leptons get the full initial-state transverse momentum
and the $p_T$ spectrum is relevant at high values;
in processes (\ref{zpc1c1}) and (\ref{c1c1dir}),
some (missing) transverse momentum is lent to
neutrinos and neutralinos, which significantly decreases the $p_T$ of $\ell^+$
and $\ell^-$.  
For direct charginos (\ref{c1c1dir})
there is no cutoff on the $\chi^+_1\chi^-_1$  
invariant mass
and the leptons are quite soft; in  $Z'_\psi\to\chi^+_1\chi^-_1$, the
leptonic $p_T$ is instead higher than in process (\ref{c1c1dir}).
\begin{figure}[htp]
\centerline{\resizebox{0.42\textwidth}{!}{\includegraphics{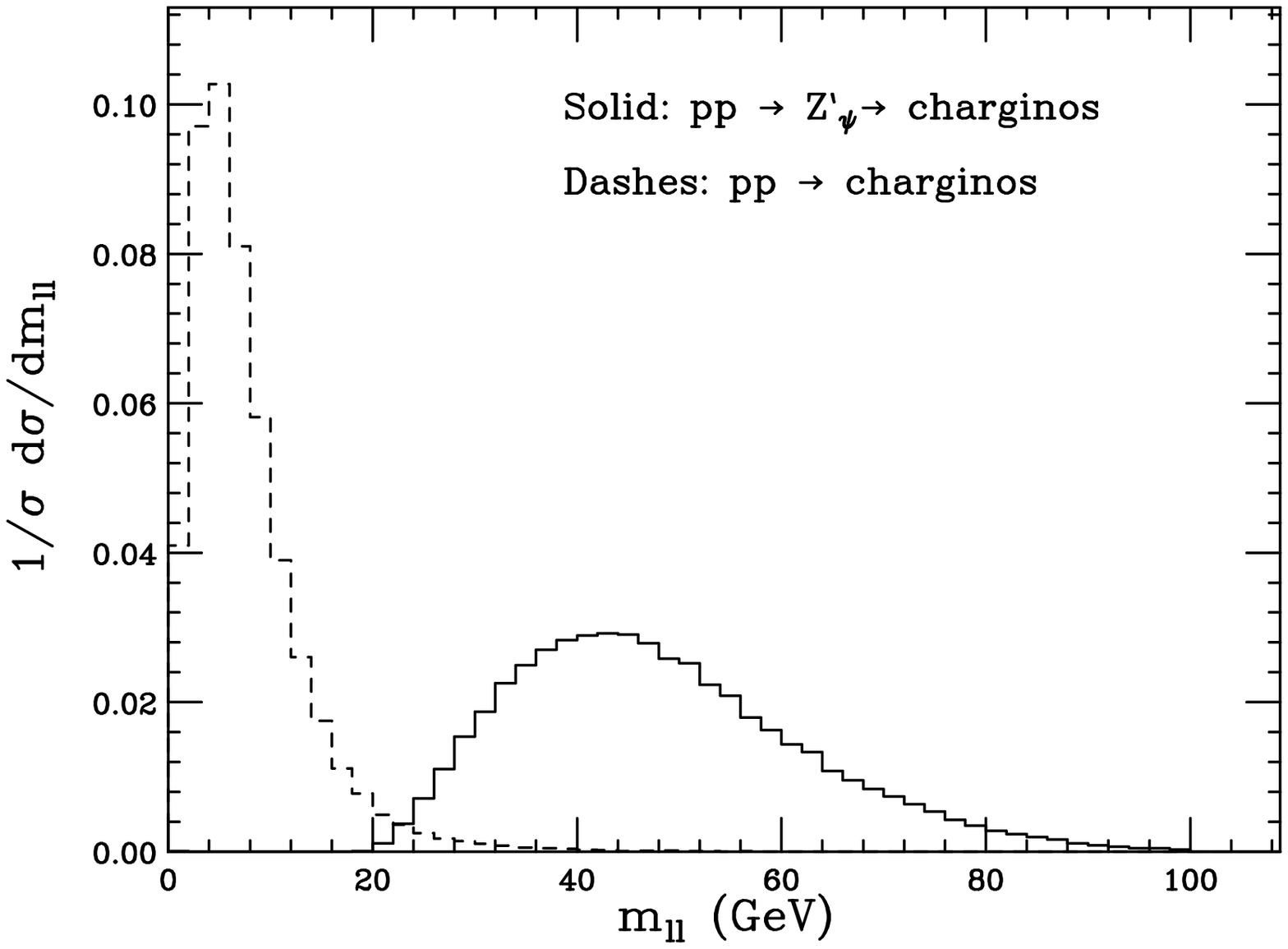}}%
\hfill%
\resizebox{0.42\textwidth}{!}{\includegraphics{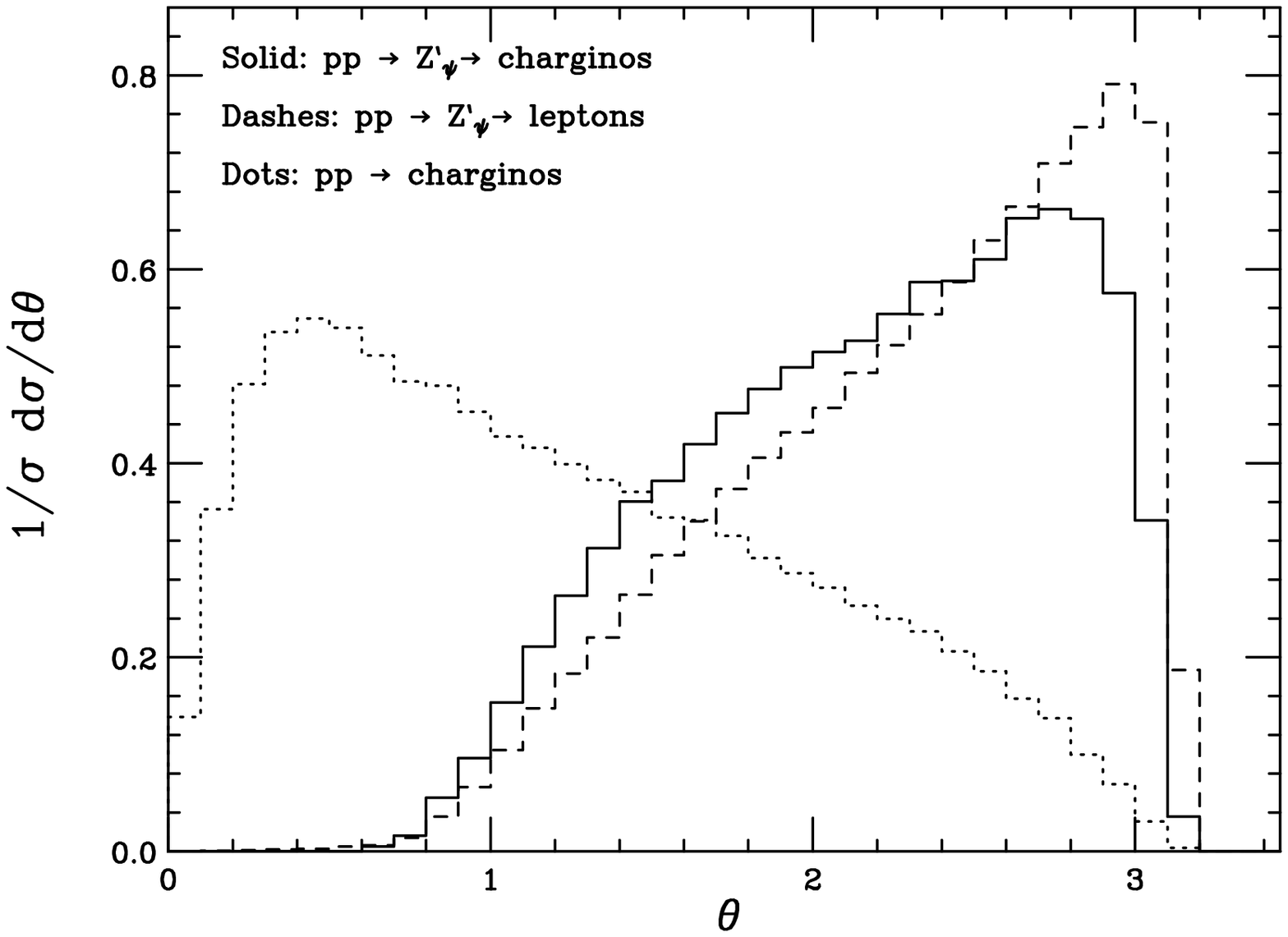}}}
\caption{Left: dilepton invariant mass, with $\protect\ell^+$ and 
$\protect\ell^-$ 
originated from $\protect Z'_\psi$ decays into charginos and from
direct $\protect pp\to \tilde\chi^+_1\tilde\chi^-_1$ events.
Right: angle between $\protect\ell^\pm$ in the laboratory frame, in all three
processes.}
\label{zpsimth}
\end{figure}
\begin{figure}[htp]
\centerline{\resizebox{0.42\textwidth}{!}
{\includegraphics{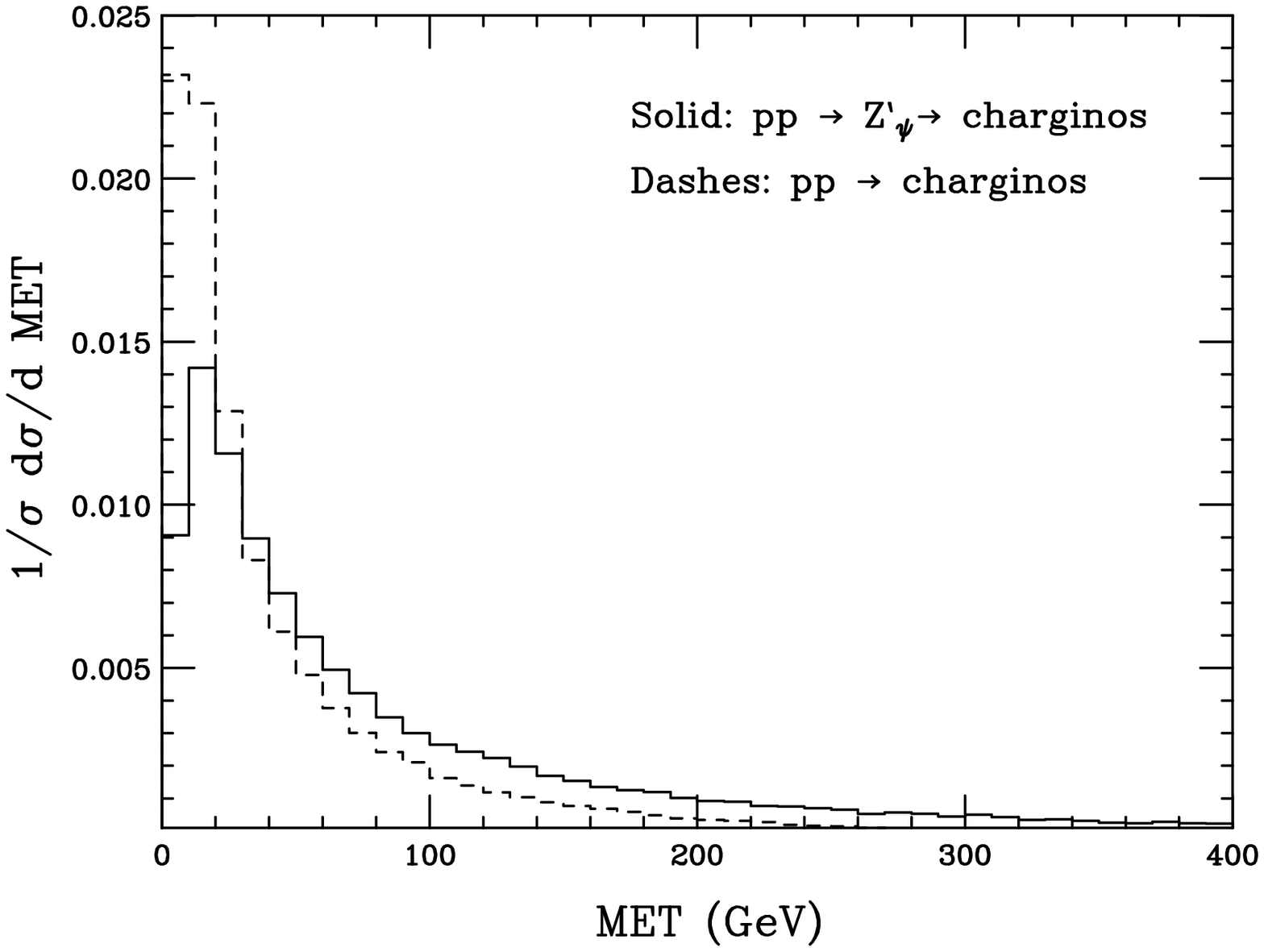}}%
\hfill%
\resizebox{0.42\textwidth}{!}{\includegraphics{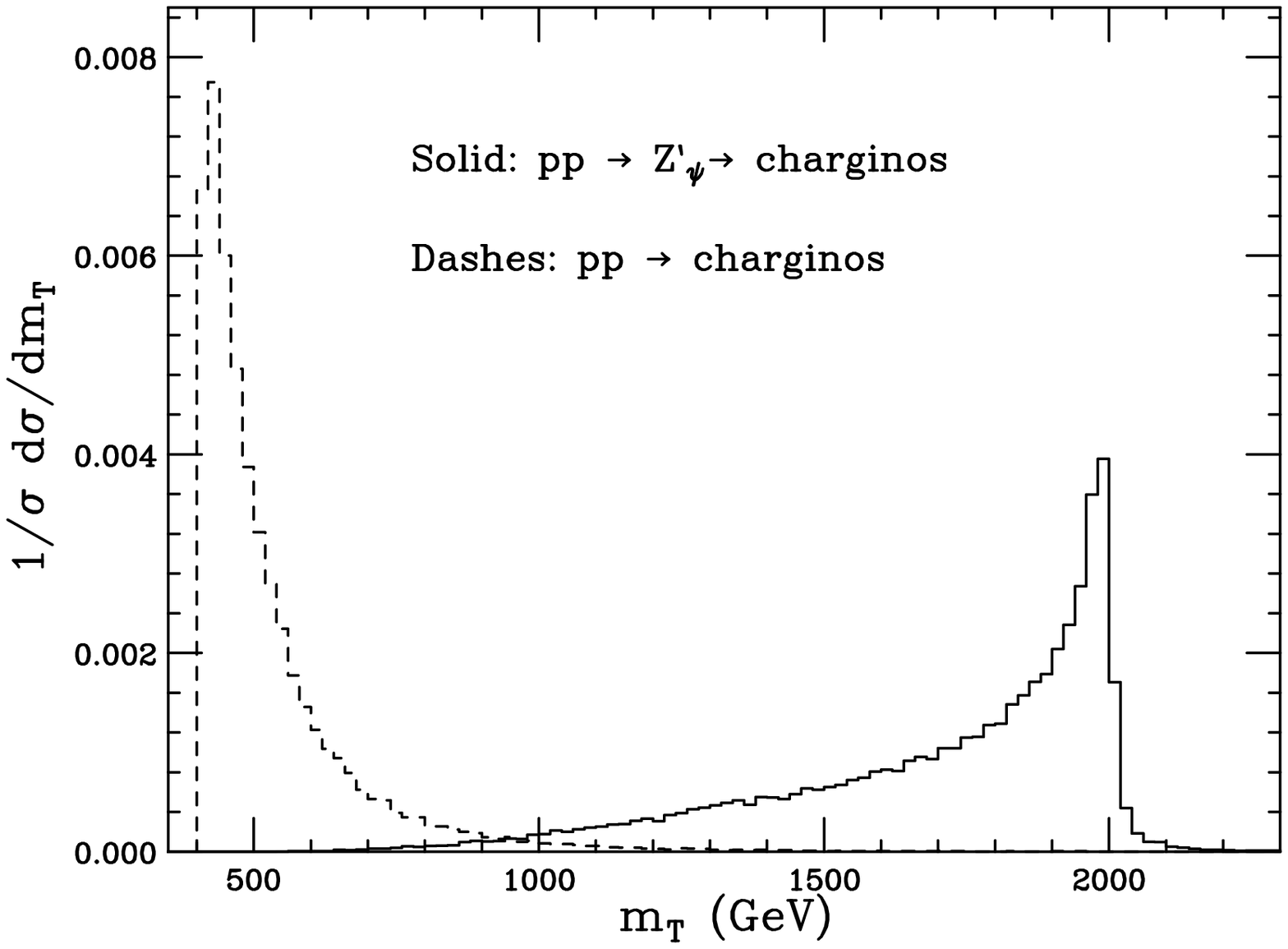}}}
\caption{
Left: missing transverse energy due to the neutrinos and neutralinos in
the cascade initiated by a primary $\protect 
Z'_\psi\to\tilde\chi^+_1\tilde\chi^-_1$ decay
(solid histogram) and for direct chargino production (dashes).
Right: transverse mass for the final-state particles 
(leptons, neutrinos and neutralinos) in the same processes.}
\label{mtmetzpsi}
\end{figure}
\par Fig.~\ref{zpsimth} presents the 
$\ell^+\ell^-$ invariant mass $m_{\ell\ell}$
(left) and the angle $\theta$ between the two charged leptons in the laboratory 
frame (right). 
In the cascade (\ref{zpc1c1}), $m_{\ell\ell}$ 
lies in the range 20 GeV~$<m_{\ell\ell}<$~100 GeV and is
maximum at $m_{\ell\ell}\simeq 45$~GeV;
for direct chargino production, $m_{\ell\ell}$ 
is peaked about 5 GeV and rapidly decreases.
As for the $\theta$ spectrum, in $Z'_\psi$ direct leptonic decays
it exhibits a maximum 
about $\theta\simeq 3$, i.e. $\ell^+$ and $\ell^-$ are almost
back-to-back;
in $Z'_\psi\to\tilde\chi^+_1\tilde\chi^-_1$, 
the $\theta$ distribution is broader and peaked
at a lower $\theta\simeq 2.75$;
in the chain (\ref{c1c1dir}), 
$\ell^+$ and $\ell^-$ are produced at smaller $\theta$.

Figure~\ref{mtmetzpsi} displays the distributions of
the sum of the transverse momenta
of `invisible' particles (neutrinos and neutralinos), 
i.e. the so-called
MET (missing transverse energy), and the transverse mass $m_T$ of
all final-state particles in Eqs.~(\ref{zpc1c1}) and (\ref{c1c1dir}).
In both processes, 
the MET spectrum is significant in the low range:
in the chain (\ref{zpc1c1}) it is 
sharply peaked at MET$\simeq 20$~GeV and smoothly decreases at larger
MET values;
for direct chargino production, the MET exhibits an even sharper
peak at MET$\simeq 10$~GeV.
As for the transverse mass,
in (\ref{c1c1dir}) it is
substantial only at small $m_T$;
in (\ref{zpc1c1}) it is instead
relevant in the range $m_{Z'}/2<m_T<m_{Z'}$
and peaked just below the $Z'_\psi$ mass threshold, 
2 TeV in the present reference point.

\begin{figure} 
\centerline{\resizebox{0.42\textwidth}{!}{\includegraphics{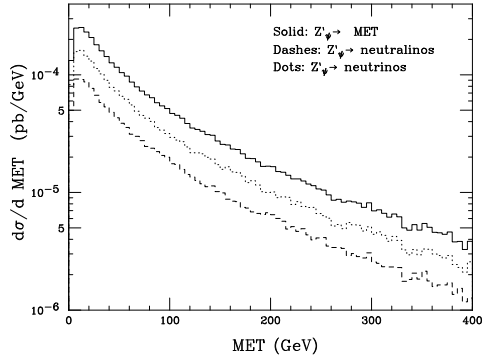}}}
\caption{MET spectrum in $\protect Z'_\psi$ decays: plotted are
the neutralino (dashes), neutrino (dots) and total (solid) contributions to 
the missing transverse energy.}
\label{metneu}
\end{figure}

Decays into neutralino pairs $\tilde\chi^0_1\tilde\chi^0_1$, accounting
for almost 5\%, are relevant for the searches
for Dark Matter candidates and yield a cross section
$\sigma(pp\to Z'_\psi\to \tilde\chi^0_1\tilde\chi^0_1)\simeq 
6.4\times 10^{-3}$~pb at 14 TeV LHC.
Therefore, about 640 events at ${\cal L}=100$~fb$^{-1}$ 
and up to almost $2\times 10^3$ at 300 fb$^{-1}$ can be foreseen, 
with typical signatures given by mono-photon or mono-jet final states.
Competing processes are $Z'_\psi$ decays into neutrino pairs,
amounting to about $\sigma(pp\to Z'_\psi\to\nu\bar\nu)\simeq 
1.1\times 10^{-2}$~pb at 14 TeV, with ${\cal O}(10^3)$ 
events at 100 and 300 fb$^{-1}$.
Figure~\ref{metneu} displays the total
missing transverse energy (MET) spectrum and the contribution due to 
neutrino and neutralino pairs in $Z'_\psi$ decays; unlike
previous distributions, they are normalized to the total LO 
cross section and not to unity, in such a way to appreciate the
discrepancy between the two subprocesses.
The shapes of all distributions are roughly the same, but
the total number of events
at any MET value is higher by about 60\%
if neutralinos contribute.
\subsection{Phenomenology - U$(1)'_\eta$ model}
Hereafter, I investigate the U(1)$'_\eta$ model, i.e.
$\theta=\arccos{\sqrt{5/8}}$ in (\ref{ztheta}),
for $m_{Z'}=2$~TeV, MSSM parameters as in
Eqs.~(\ref{mubeta}) and (\ref{aaa}) and soft sfermion masses 
fixed to the following values:
\begin{equation}
m_{\tilde\ell}^0=m_{\tilde\nu_\ell}^0=1.5~{\rm TeV},\ 
m^0_{\tilde q}=3~{\rm TeV}.
\end{equation}
After adding D- and F-terms, the masses of squarks and sleptons are
those quoted in Tables~\ref{tabmassqeta} and \ref{tabmassleta}, whereas
Table~\ref{tabmassheta}  and 
\ref{tabmasscneta} contain the masses of Higgs bosons and gauginos,
respectively.
\begin{table}[htp]
\caption{Squark masses in GeV in the 
$\protect Z'_\eta$ model.}
\label{tabmassqeta}
\begin{center}
\small
\begin{tabular}{|c|c|c|c|c|c|}
\hline
$m_{\tilde d_1}$ &   $m_{\tilde u_1}$ &  $m_{\tilde s_1}$ &  $m_{\tilde c_1}$ &  
$m_{\tilde b_1}$ & $m_{\tilde t_1}$\\ 
\hline
3130.8  & 3129.8  & 3130.8 & 3129.8 & 3130.8 & 3175.5 \\
\hline
$m_{\tilde d_2}$ &   $m_{\tilde u_2}$ &  $m_{\tilde s_2}$ &  $m_{\tilde c_2}$ &  
$m_{\tilde b_2}$ & $m_{\tilde t_2}$\\ 
\hline
 3065.9 & 2863.6  & 3065.9 & 2863.6 & 3065.9 &  2823.5 \\
\hline\end{tabular}
\end{center}
\end{table}
\begin{table}[htp]
\caption{Slepton masses in the $\protect Z'_\eta$ scenario.}
\label{tabmassleta}
\begin{center}
\small
\begin{tabular}{|c|c|c|c|c|c|c|c|}
\hline
$m_{\tilde \ell_1}$ &   $m_{\tilde \ell_2}$ &  $m_{\tilde\tau_1}$
& $m_{\tilde\tau_2}$ &
$m_{\tilde \nu_{\ell,1}}$ &  $m_{\tilde \nu_{\ell,2}}$ &
$m_{\tilde \nu_{\tau,1}}$ &  $m_{\tilde \nu_{\tau,2}}$ \\ 
\hline
 1194.6 & 1364.5  & 1208.8 & 1307.7 &  1361.8  & 456.0 & 1368.0 & 456.0\\ 
\hline\end{tabular}
\end{center}
\end{table}
\begin{table}[htp]
\caption{Higgs bosons in the $Z'_\eta$ model, with masses expressed in GeV.}
\label{tabmassheta}
\begin{center}
\small
\begin{tabular}{|c|c|c|c|c|}
\hline
$m_h$ &   $m_H$ &  $m_{H'}$ &  $m_A$ & $m_{H^+}$\\ 
\hline
  124.9 &  2004.2 & 4229.4  & 4229.4 & 4230.0\\ 
\hline\end{tabular}
\end{center}
\end{table}
\begin{table}[htp]
\caption{Masses in GeV of charginos and neutralinos in the $Z'_\eta$ model. }
\label{tabmasscneta}
\begin{center}
\small
\begin{tabular}{|c|c|c|c|c|c|c|c|}
\hline
$m_{\tilde\chi_1^+}$ &   $m_{\tilde\chi_2^+}$ & $m_{\tilde\chi_1^0}$ &   $m_{\tilde\chi_2^0}$ 
& $m_{\tilde\chi_3^0}$ &   $m_{\tilde\chi_4^0}$ 
& $m_{\tilde\chi_5^0}$ &   $m_{\tilde\chi_6^0}$  \\ 
\hline
 206.5 & 882.4 & 199.3  & 212.5 & 408.2 & 882.3 & 1562.8 & 2569.2 \\ 
\hline\end{tabular}
\end{center}
\end{table}
\par Table~\ref{tabbreta} presents the branching ratios of the 
$\protect Z'_\eta$: within supersymmetry,
sneutrino pairs $\protect\tilde\nu_2\tilde\nu_2^*$ exhibit the highest rate,
slightly below 10\%. It is therefore worthwile investigating the cascade:
\begin{equation}
pp\to Z'_\eta\to \tilde\nu_2\tilde\nu_2^*\to (\tilde\chi^0_2\nu_2) (\tilde\chi^0_2\bar\nu_2)
\to (\ell^+\ell^-\tilde\chi^0_1\nu_2)(\ell^+\ell^-\tilde\chi^0_1\bar\nu_2).
\label{zpsnu}
\end{equation}
The branching ratios of 
sneutrinos and neutralinos are given in Tables~\ref{brsnu}
and \ref{brchi02}.
\begin{table}[t] 
\caption{$\protect Z'_\eta$ decay rates in the MSSM reference point for
a mass $\protect m_{Z'}=2$~TeV.}
\label{tabbreta}
\begin{center}
\small
\begin{tabular}{|c|c|}
\hline
Final State & $Z'_\eta$ Branching ratio (\%) \\
\hline
$\tilde\chi_1^+\chi_1^-$ & 5.6 \\
\hline
$\tilde\chi_1^0\tilde\chi_1^0$ & 1.9 \\
\hline
$\tilde\chi_2^0\tilde\chi_2^0$ & 2.1 \\
\hline
$\tilde\chi_1^0\tilde\chi_2^0$ & 1.5 \\
\hline
$\sum_\ell\tilde\nu_{\ell,2}\tilde\nu_{\ell,2}^*$ & 9.4  \\
\hline
$hZ$ & 1.5\\
\hline
$W^+W^-$ & 3.0\\
\hline
$\sum_i d_i\bar d_i$ & 16.1 \\
\hline
$\sum_i u_i\bar u_i$ & 25.5 \\
\hline
$\sum_i \nu_i\bar \nu_i$ & 27.8 \\
\hline
$\sum_i \ell^+_i\ell^-_i$ & 5.3 \\
\hline
\end{tabular}
\end{center}
\end{table}
\begin{table} 
\caption{Sneutrino $\protect \tilde\nu_2$ branching ratios, in the 
representative
point of the $\protect Z'_\eta$ model.}
\label{brsnu}
\begin{center}
\small
\begin{tabular}{|c|c|}
\hline
Final state & $\tilde\nu_2$ branching ratio (\%) \\
\hline
$\tilde\chi_1^0\nu_2$ & 4.0 \\
\hline
$\tilde\chi_2^0\nu_2$ & 37.3 \\
\hline
$\tilde\chi_3^0\nu_2$ & 58.7 \\
\hline\end{tabular}
\end{center}
\end{table}
\begin{table} 
\caption{Decay rates for the lighter neutralino
$\protect \tilde\chi^0_2$.}
\label{brchi02}
\begin{center}
\small
\begin{tabular}{|c|c|}
\hline
Final State & $\tilde\chi^0_2$ Branching ratio (\%) \\
\hline
$\sum_i\tilde\chi_1^0q_i\bar q_i$ & 63.3 \\
\hline
$\sum_i\tilde\chi^0_1\ell^+_i\ell^-_i$ & 13.4 \\
\hline
$\sum_i\tilde\chi_1^0\nu_i\bar\nu_i$ & 20.6 \\
\hline\end{tabular}
\end{center}
\end{table}
The cross section of the cascade (\ref{zpsnu}) is 
$\sigma(pp\to Z'_\eta\to 4\ell+{\rm MET})\simeq 1.90\times 10^{-4}$~pb
\cite{cor}.
\begin{figure}
\centerline{\resizebox{0.42\textwidth}{!}
{\includegraphics{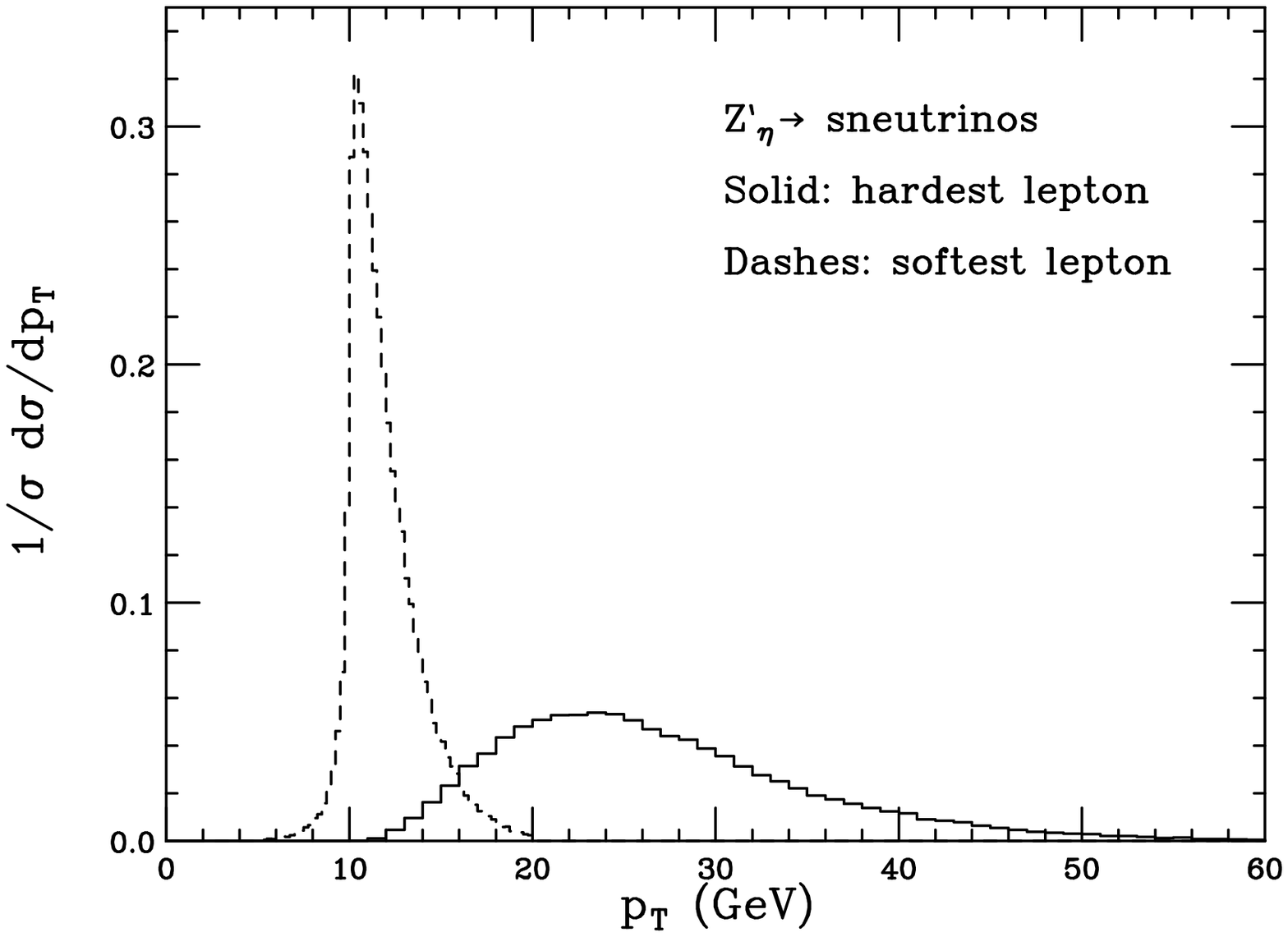}}%
\hfill%
\resizebox{0.42\textwidth}{!}{\includegraphics{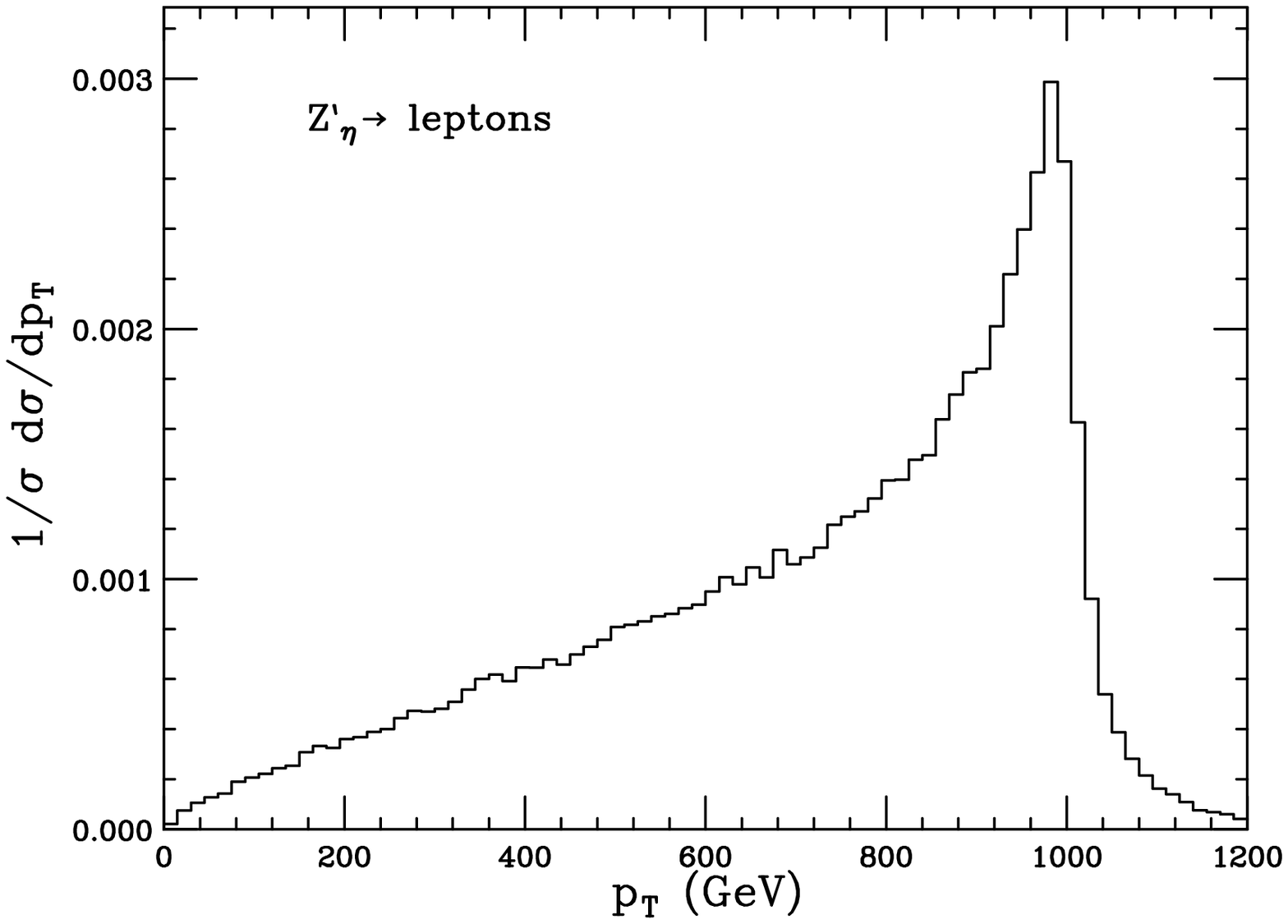}}}
\caption{Left: Transverse momentum of the hardest (solid) and softest (dashes) lepton
in the cascade initiated by a primary decay into sneutrinos. Right: Lepton transverse momentum in 
$\protect Z'_\eta\to \ell^+\ell^-$ processes.}
\label{zetapt}
\end{figure}
\par In Fig.~\ref{zetapt} the 
transverse momenta of $\ell^\pm$ in direct decays $Z'_\eta\to\ell^+\ell^-$ 
and of the softest and hardest lepton 
in the decay chain (\ref{zpsnu}) are
plotted.
In the cascade, the hardest lepton has a broad spectrum, relevant
in the 10 GeV$<p_T<50$~GeV range, whereas
the $p_T$ of the softest $\ell^\pm$ is a narrow distribution, 
substantial only for 8 GeV$<p_T<20$~GeV;
the spectrum in the direct production $Z'_\eta\to\ell^+\ell^-$ 
is roughly the same as in the $Z'_\psi$ case.
\begin{figure}
\centerline{\resizebox{0.42\textwidth}{!}
{\includegraphics{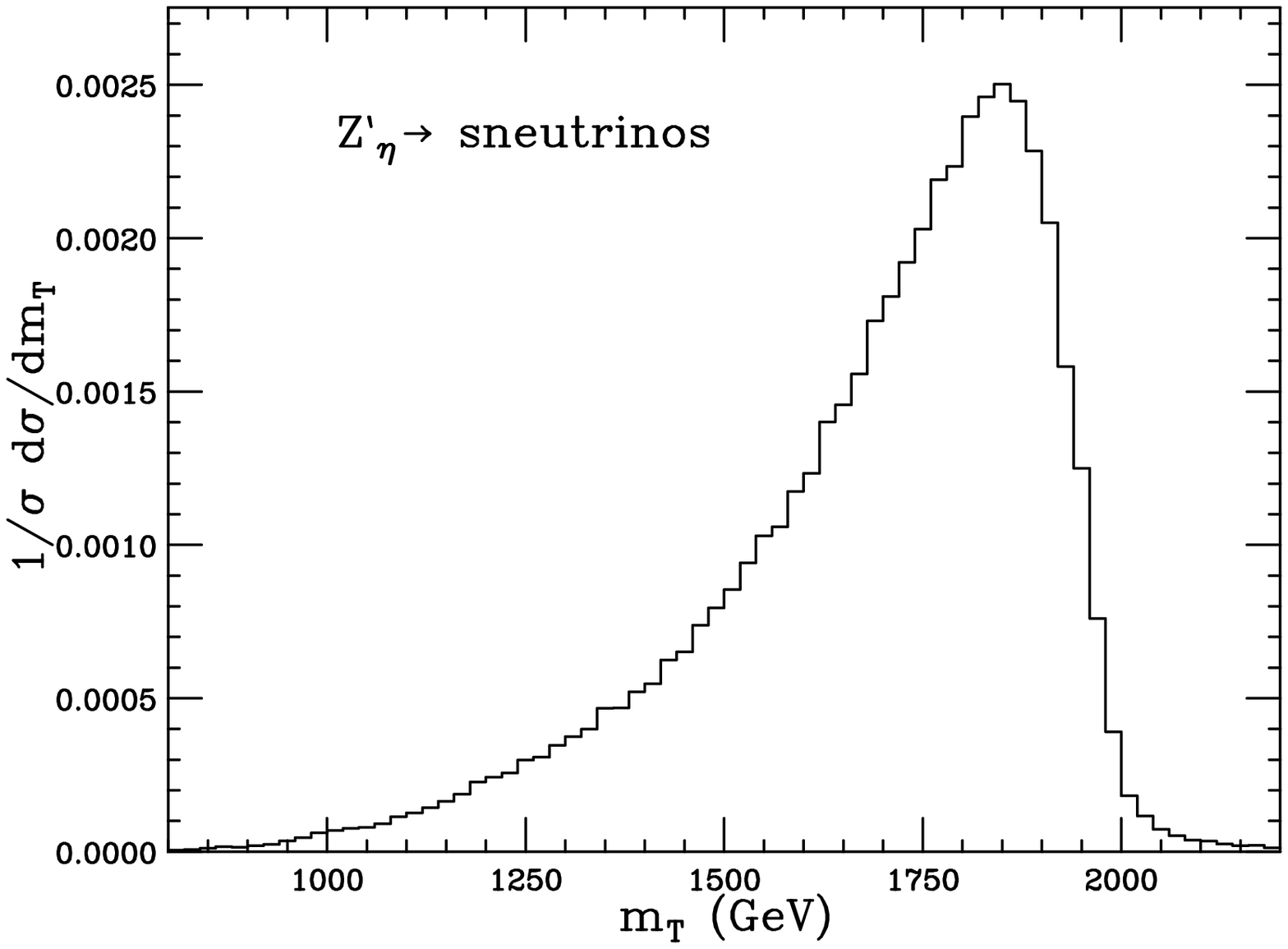}}%
\hfill%
\resizebox{0.42\textwidth}{!}{\includegraphics{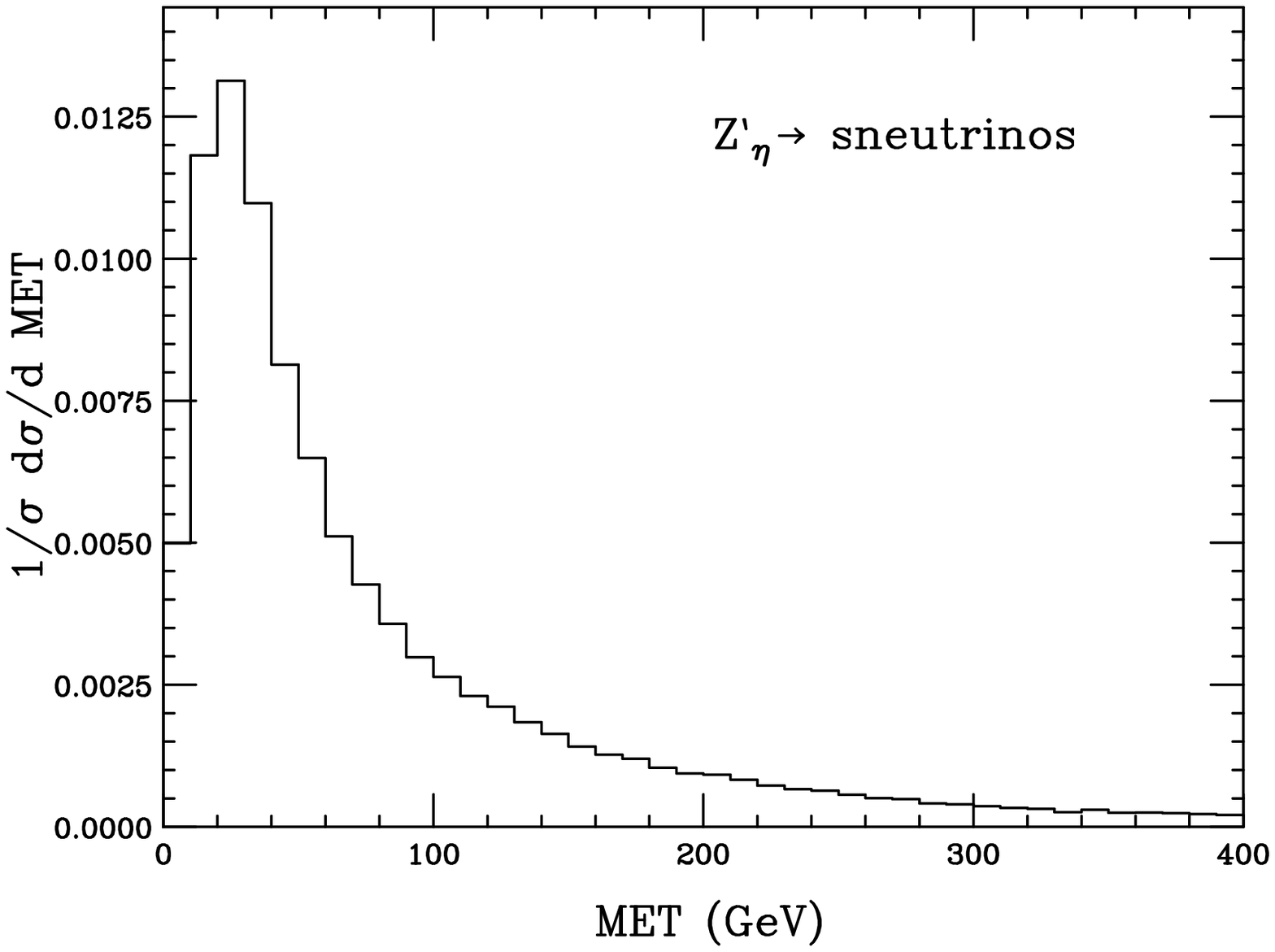}}}
\caption{Left: transverse mass of final states initiated
by a decay of the $\protect Z'_\eta$ into sneutrino pairs
and leading to a final state with four leptons and MET.
Right: missing transverse energy due to neutrinos and neutralinos.}
\label{mtmetzeta}
\end{figure}\par
Figure~\ref{mtmetzeta} presents the 
spectrum of the missing transverse energy
and of the transverse mass of the final state (\ref{zpsnu}):
the MET distribution is similar to the $Z'_\psi$ one; the transverse mass
is relevant in the range $m_{Z'}/2<m_T<m_{Z'}$ and is overall a broader
and smoother distribution.

\newpage\section{Conclusions}
I investigated supersymmetric $Z'$ decays at the LHC,
for $\sqrt{s}=14$~TeV, in GUT-inspired U(1)$'_\psi$ and U(1)$'_\eta$
gauge models.
The analysis was carried out for few reference points of the
MSSM, extended to account for the new 
U(1)$'$ symmetry, and consistent with a 125 GeV
Higgs boson.
Both  $Z'_\psi$ and $Z'_\eta$ models
yield substantial supersymmetric event rates in the
14 TeV run of the LHC: $Z'$ decays
into chargino, sneutrino or neutralino pairs lead to final
states with leptons and missing energy, which
can be discriminated from  
$Z'\to\ell^+\ell^-$ processes and from
direct sparticle production.
Once data on high-mass leptons at 14~TeV
are available, it will be interesting comparing them with
the theory predictions,
as done in \cite{cor1} at 8 TeV,
and possibly determine the exclusion limits accounting for
supersymmetry. 
Nevertheless, a full analysis should compare such 
signals with the backgrounds due to 
SM, other supersymmetric processes and
 non-supersymmetric $Z'$ decays, and account for the detector simulation. 
The inclusion of background  
and detector effects is in progress.

\end{document}